# Surface lattice kink solitons

Yaroslav V. Kartashov,[1] Victor A. Vysloukh,[2] and Lluis Torner[1]

[1]*ICFO-Institut de Ciencies Fotoniques, and Universitat Politecnica de Catalunya, Mediterranean Technology Park, 08860 Castelldefels* (*Barcelona*), *Spain*
[2]*Departamento de Fisica y Matematicas, Universidad de las Americas – Puebla, Santa Catarina Martir, 72820, Puebla, Mexico*
*Yaroslav.Kartashov@icfo.es*

**Abstract:** We predict theoretically that surface of an optical lattice imprinted in defocusing nonlinear media can support shock, or kink waves. Such new surface waves contain a modulationally stable pedestal and are strongly localized at the edge of the optical lattice due to Bragg-type reflection. The kink steepness and localization degree can be controlled by the lattice depth. We found two types of kinks, which exhibit distinct stability properties for each finite gap in the lattice spectrum. Our findings open the way to experimental observation of optical surface kink waves.

Kinks are very specific type of nonlinear waves. They resemble shock waves having different amplitude asymptotes far from the wave center where they exhibit a sharp transition region. In fluid mechanics and in gas dynamics shock waves typically occur when the velocity of the nonlinear excitation depends on its amplitude. Different types of shock waves were encountered in plasma physics and in solid-state physics [1,2].

Kinks may form with light [3,4] and this process is accompanied by remarkable nonlinear spectral broadening associated to the sharp wavefront. A tendency to sharp wavefront formation (or self-steepening) in *time domain* is common for the high-intensity short light pulses in nonlinear fibers with normal group-velocity dispersion [5] or in optical capillaries filled with organic liquids with delayed nonlinear response [6]. The inertial Raman response and higher-order group velocity dispersion also lead to self-steepening [7]. However, in these settings shock waves exist only as transient objects because of lack of broadband phase matching of wave's spectral components required for existence of stationary solutions.

The *space domain* offers richer possibilities for stationary shock wave formation, due to the possibility to control the diffraction properties of beams (i.e., phasing of spatial spectral components) in fabricated [8] or optically-induced [9] periodic lattices. Photorefractive materials are excellent candidates for experiments with spatial shock waves since they offer tunability of nonlinear response and might be used for lattice induction [10-13]. Bulk photorefractive crystals with local drift and nonlocal diffusion nonlinearities allow formation of steady-state shock waves, which, however, contain modulationally unstable pedestals [14]. Another mechanism of shock waves formation based on two-wave mixing was explored in [15,16]. Extended soliton trains in lattices with saturable [17] or quadratic [18] nonlinearity may also be viewed as kinks.

Recently, the concept of surface waves existing at the interface of periodic and uniform medium was suggested [19]. Due to shallow refractive index modulations in periodic medium that can be created with currently available technologies surface waves at lattice interfaces can be observed at achievable power levels [20], in contrast to nonlinear surface waves at the interfaces of natural materials [21-23]. Lattice interfaces support localized gap surface solitons [24-26]. Still open and challenging problem is the existence and stability of surface kinks at the interface of a bulk medium and an optical lattice. In this paper we introduce such surface shock waves, consisting of the constant stable pedestal located in a uniform self-defocusing medium and of spatially fading amplitude oscillations inside the lattice. These are *proper kink waves*, i.e., waves featuring a transition between regions with nonzero and zero light intensities, that do not exist at interfaces of two uniform materials with different linear refractive indices and equal nonlinear coefficients. They can only exist when both linear and nonlinear properties of two uniform materials are different [27,28] or in suitable nonlinear waveguides [29], under conditions not easy to meet in practice. In contrast, here we address surface kinks in a single nonlinear material. The steepness of surface lattice kinks, their internal structure, and penetration depth into periodic structure can be controlled by varying the lattice depth and its period. We found two types of surface shock waves featuring distinct stability properties: an unstable branch that transforms into the family of known localized gap solitons, and the important branch of *stable kink waves,* which can not be asymptotically derived from any family of localized solitons found before, and that thus provides an experimentally feasible setting for the observation of kink surface waves at moderate power levels in contrast to all kinks discussed earlier.

We thus address the propagation of a laser beam at the edge of a semi-infinite lattice imprinted in a defocusing cubic Kerr-type medium, described by the nonlinear Schrödinger equation for the dimensionless complex amplitude of the light field $q$:

$$i\frac{\partial q}{\partial \xi} = -\frac{1}{2}\frac{\partial^2 q}{\partial \eta^2} + q|q|^2 - pR(\eta)q. \tag{1}$$

In Eq. (1) the transverse $\eta$ and longitudinal $\xi$ coordinates, and field amplitude are expressed in soliton units [7]. The parameter $p$ describes the lattice depth, while the function $R(\eta) \equiv 0$ for $\eta < 0$ and $R(\eta) = 1 - \cos(\Omega\eta)$ for $\eta \geq 0$ stands for the profile of transverse refractive index modulation, with $\Omega$ being the frequency of modulation. The material nonlinearity is uniform in the medium. The presence of the lattice results in an increase of the mean effective refractive index in the region $\eta \geq 0$. We assume that the depth of the linear refractive index modulation is small compared with the unperturbed index and is of the order of contribution arising via the defocusing nonlinearity. Such interfaces can be realized, e.g., in AlGaAs [30] or LiNbO [31], by etching of periodic structure on top of a suitable substrate. The technique of optical induction [10-13], combined with erasing parts of the red-light-imprinted lattice with an intense green background illumination, may provide an alternative tool for realization of interfaces considered here. In both cases one can achieve sufficiently sharp transition between the lattice and the uniform medium. Here we concentrate on the case of defocusing media where modulational instabilities are suppressed. Based on the scaling properties of Eq. (1) that can be used to obtain different families of solutions, here we set $\Omega = 4$.

We address kink solutions of Eq. (1) that contain a constant pedestal inside the uniform medium ($\eta < 0$) and gradually decaying tail inside the lattice ($\eta \geq 0$). Wave localization inside the periodic structure imposes restrictions on the wave parameters. To elucidate them it is important to consider the Floquet-Bloch spectrum of the lattice. We thus search for Bloch modes of the infinite periodic lattice $R(\eta) = 1 - \cos(\Omega\eta)$ for $\eta \in (-\infty, +\infty)$ in the form $q(\eta, \xi) = w(\eta)\exp(ib\xi + ik\eta)$, where $b$ is a real propagation constant, $k$ is a transverse Bloch wavenumber, and $w(\eta) = w(\eta + 2\pi/\Omega)$ is a complex periodic function. Upon substitution into the linear version of Eq. (1), one arrives at the eigenvalue problem

$$bw = \frac{1}{2}\left(\frac{d^2w}{d\eta^2} + 2ik\frac{dw}{d\eta} - k^2w\right) + pRw \tag{2}$$

that can be solved numerically to obtain the dependencies $b(k,p)$. For each lattice depth all possible propagation constants of the linear Bloch modes are joined into bands (Fig. 1(a)), where no localized solutions exist even in the nonlinear regime. Localized solitons emerge as nonlinear defect modes in the gaps of Floquet-Bloch spectrum. Such requirement also holds for any type of nonlinear wave supported by the lattice interface, which exhibits vanishing tails inside the lattice. Notice that Floquet-Bloch spectrum possesses a single semi-infinite gap and infinite number of finite gaps. Because of the defocusing nonlinearity, only finite gaps can give rise to soliton solutions.

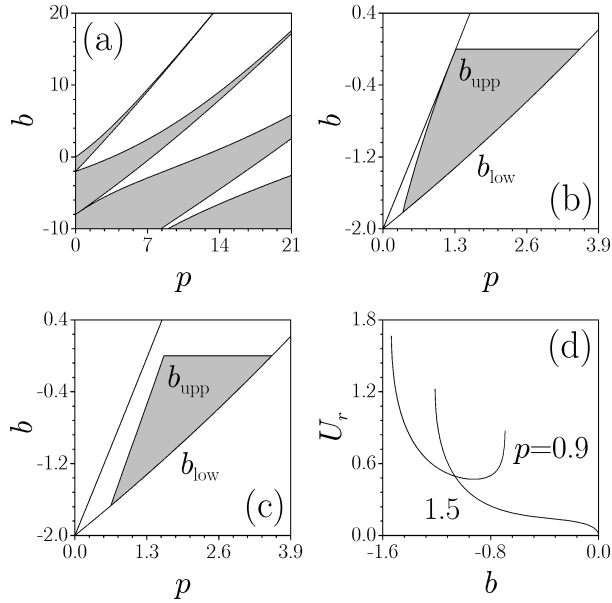

Fig. 1. (a) Bang-gap structure of periodic lattice. Bands are shown gray, gaps are shown white. Domains of existence of "out-of-phase" (b) and "in-phase" (c) shock surface waves. (d) Reduced energy versus propagation constant for "out-of-phase" shock waves. In all cases $\Omega = 4$.

In the case of a nonlinear interface between a lattice and a uniform medium the steady-state kink surface waves have the form $q(\eta,\xi) = w(\eta)\exp(ib\xi)$, where $w(\eta)$ is real function. We found the profiles of such waves numerically from Eq. (1) by using a relaxation method. Representative examples are shown in Fig. 2. In all cases the intensity of the wave drops off from a constant value at $\eta \to -\infty$ to zero at $\eta \to +\infty$. It follows from Eq. (1) that $w(\eta \to -\infty) = (-b)^{1/2}$, which means that kink waves can be found only at $b < 0$, in contrast to localized gap surface solitons requiring $b > 0$ [24]. Inside the uniform medium ($\eta \leq 0$) the profiles of the surface kinks resemble those of dark solitons, while they are localized inside the lattice due to Bragg reflection from the periodic medium. The gap-type structure of the waves inside the lattice gives rise to multiple amplitude oscillations on their wing at $\eta \geq 0$, so that the field inside the lattice features rich internal structure determined by the gap number (i.e. in different gaps the energy can be localized either in the vicinity of lattice maxima, or in between lattice maxima). This is in contrast to kink waves at uniform interfaces that always have monotonic tails and much simpler structure of existence domains. Bragg backward reflection from the periodic medium is crucial for existence of the kink waves, as confirmed by the fact that they do not exist in a system with uniform nonlinearity and lattice replaced by the equivalent constant step in refractive index. Importantly, in the absence of optical lattice surface waves whose intensity vanishes at $\eta \to +\infty$ and remains nonzero at $\eta \to -\infty$ can only be found at the interfaces of two distinct nonlinear materials [27] or in defocusing waveguides [29] and would require higher intensity levels. One of the central results of this paper is that the interfaces with lattices having shallow refractive index modulations allow formation of such "true" kink waves inside single nonlinear material for the moderate intensity levels that are of the same order as intensities required for lattice soliton formation.

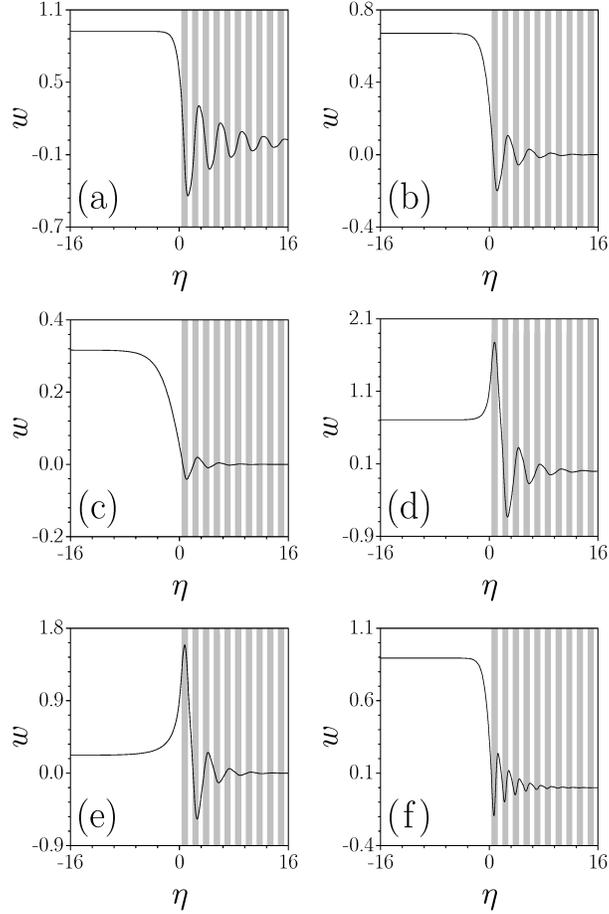

Fig. 2. Profiles of surface shock waves at (a) $b = -0.85$, (b) $b = -0.45$, (c) $b = -0.1$, (d) $b = -0.5$, (e) $b = -0.05$, and (f) $b = -0.8$. Lattice depth $p = 9$ for panel (f), while in all other panels $p = 2$. White regions correspond to $R(\eta) \leq 1$, while in gray regions $R(\eta) > 1$. In all cases $\Omega = 4$.

We found two different types of kink surface waves originating from each finite gap of the lattice spectrum. For concreteness, here we are interested only in solitons whose primary maxima are located in the nearest-to-interface lattice channel. The primary maximum of "in-phase" kinks is in-phase with the constant pedestal (Figs. 2(d) and 2(e)), while it is out-of-phase with pedestal for the "out-of-phase" kinks depicted in Figs. 2(a)-2(c). Consequently, the intensity of out-of-phase solutions inside the lattice does not exceed the limiting value $-b$, while for in-phase waves the maximum of intensity is achieved in the first lattice channel. Importantly, the central point of the wavefront resides in the first lattice channel for out-of-phase kinks while it resides in the second channel for in-phase waves.

The constraint $b < 0$ together with the structure of band-gap lattice spectrum determines the domains of existence for kink surface waves. The difference in internal structure of in-phase and out-of-phase waves manifests itself in different domains of existence (compare shaded regions in Figs. 1(b) and 1(c)). Notice that the domain of existence of kinks does not occupy a whole gap. For waves of both types the domain of existence shrinks when the lattice depth reaches a critical value (for first-gap solitons, one has $p_{\mathrm{cr}} \approx 3.56$) corresponding to conditions where the lower gap edge crosses the line $b = 0$. The domain of existence of out-of-phase waves broadens with decrease of $p$ until the value $p \approx 1.32$ is reached, where upper gap edge crosses the line $b = 0$. For $1.32 < p < 3.56$ "out-of-phase" waves *vanish* at the upper cutoff $b_{\mathrm{upp}} = 0$, since the amplitude of both, the pedestal and the wave tail inside the lattice simultaneously decreases as $b \to 0$ (Fig. 2(c)). The amplitude of the pedestal is largest near the lower cutoff $b_{\mathrm{low}}$ coinciding with the lower gap edge. At $b \to b_{\mathrm{low}}$ kinks deeply penetrate into the lattice (Fig. 2(a)). The steepness of the waves increases as $b \to b_{\mathrm{low}}$. Their penetration depth into the lattice region near upper cutoff depends on the relative position of point $b = 0$ inside the gap. As $p \to p_{\mathrm{cr}}$ the kink is weakly localized inside the lattice in the entire existence domain. For $p < 1.32$ the upper cutoff for out-of-phase wave departs from the upper gap edge. This entirely surface effect results in complete shrinking of shock wave existence domain at $p \approx 0.36$. Physically, this occurs because Bragg backward reflection can not compensate the energy flow into the lattice region that increases progressively with growth of wavefront steepness.

To fully characterize the shock waves, let us introduce the renormalized energy flow:

$$U_r = \int_{-\infty}^{\infty} [w(\eta) - |b|^{1/2} H(\eta)]^2 d\eta \qquad (3)$$

where $H(\eta) = 1$ for $\eta \leq 0$ and $H(\eta) = 0$ for $\eta > 0$. This quantity allows identifying the energy flow concentrated within the shock wave front and its tail. One can see from Fig. 1(d) that $U_r$ vanishes in the upper cutoff for $p \in [1.32, 3.56]$, while at $p \in [0.36, 1.32)$ the derivative $dU_r/db$ becomes negative at $b \to b_{\mathrm{upp}}$. It should be pointed out that at fixed $U_r$ or $b$ the shock wave steepness and its penetration depth into the lattice region can be controlled by varying the depth and the frequency modulation of the lattice. Thus, at fixed $b$, increasing the lattice depth $p$ results in a growing steepness of the wave and in an increase of the wave penetration depth into the lattice. This phenomenon is linked to the fact that one approaches the lower edge of the kink existence domain, where the wave located inside the lattice closely resembles the profile of a Bloch wave. A similar scenario is encountered when decreasing the lattice modulation frequency for fixed $b$ and $p$. The important new physical properties afforded by the lattice are thus clearly apparent.

The domain of existence of in-phase shock wave first expands with decrease of $p$ down to 1.61 (notice that this value exactly coincides with critical lattice depth for existence of localized surface gap solitons [24]), but then gets narrower and shrinks at $p \approx 0.65$ (Fig. 1(c)). Therefore, the existence domain for in-phase waves is smaller than that for out-of-phase waves. The key difference between these two types of waves is that for $p \in [1.61, 3.56]$ and at $b \to b_{\mathrm{upp}} = 0$ the in-phase wave transforms into fully localized surface gap soliton, and can be viewed as a continuation of family of gap surface solitons in the region $b < 0$, in contrast to the entirely new family of out-of-phase waves that can not be asymptotically derived from any localized gap surface soliton family found earlier, because such solitons do not transform into localized waves but vanish at the cutoff, thus remaining spatially extended. The domain of existence for in-phase waves is adjacent to the domain of existence of gap solitons at $b > 0$. The height of the pedestal of in-phase shock wave gradually decreases as $b \to b_{\text{upp}}$ (compare Figs. 2(d) and 2(e)), while wave tail inside the lattice closely resembles gap soliton profile at $p \in [1.61, 3.56]$. Irrespectively of the value of $p$, the renormalized energy flow acquires its minimal value deep inside existence domain, while the derivative $dU_r/db < 0$ near both lower and upper cutoffs.

A similar picture was encountered for all finite gaps of the lattice spectrum, i.e. each gap gives rise to two types of shock waves. An example of out-of-phase shock wave originating from the second finite gap is shown in Fig. 2(f). The domain of existence of such solutions in terms of lattice depth $p$ increases with growing number of the gap.

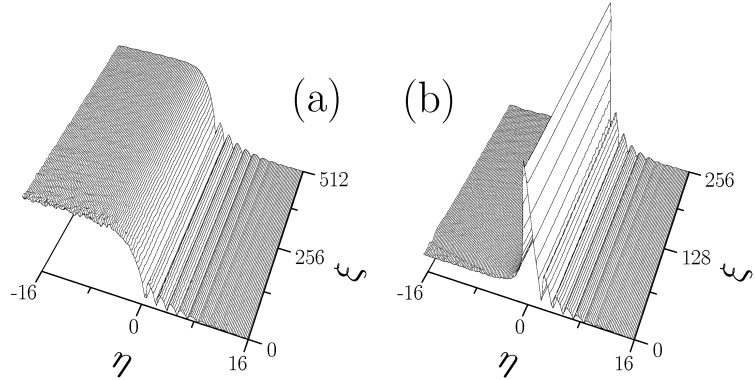

Fig. 3. Propagation of perturbed (a) "out-of-phase" shock surface wave at $b = -0.2$, $p = 3$, and (b) "in-phase" wave at $b = -0.1$, $p = 3$. In all cases $\Omega = 4$.

One of the central results of this paper is that certain types of surface kink waves can be *completely stable*. To elucidate the stability of the obtained solutions we performed extensive simulations of Eq. (1) with the input conditions $q|_{\xi=0} = w(1+\rho)$, where $\rho(\eta)$ stands for the broadband noise with Gaussian distribution and variance $\sigma^2_{\text{noise}} = 0.01$. In defocusing media the constant pedestal is stable, so that the only type of instability that may develop is an oscillatory instability typical for gap solitons. We found that out-of-phase waves are *completely stable* in a substantial part of their existence domain. They are stable everywhere, except for a narrow region near the lower cutoff where oscillatory instability may take place. Such waves keep their input structure on propagation over indefinitely long distances in the presence of considerable input noise (Fig. 3(a)). At the same time in-phase waves were found to be always *unstable*. The oscillatory instabilities of such waves develop mainly on their tails inside the lattice and typically are so weak that waves can propagate without any appreciable distortion over hundreds of diffraction lengths (Fig. 3(b)). Surprisingly, out-of-phase wave originating from the second gap (Fig. 2(e)) can also be stable despite the fact that gap soliton with similar symmetry in infinite lattice typically undergo strong exponential instability.

In summary, we predicted the existence of kink or shock surface waves at the interface of optical lattices imprinted in defocusing nonlinear media. To the best of our knowledge, such waves provide the first known example of completely stable surface kink waves in optics. In addition, our findings open the way to the first experimental observation of surface kinks, a goal not yet achieved to date because of lack of suitable physical setting. The kinks predicted here can be excited, e.g., in $\text{LiNbO}_3$ waveguide arrays exhibiting required defocusing photo-voltaic nonlinearities at low intensity levels, similar to those required in recent experiments on surface solitons [25,26].

**Acknowledgements**

This work has been supported in part by the Government of Spain through the Ramon-y-Cajal program and through the grant TEC2005-07815/MIC.